\documentclass{vgtc}                 

\graphicspath{{figures/}{pictures/}{images/}{./}} 

\usepackage{times}                     

\usepackage{tabu}                      
\usepackage{booktabs}                  
\usepackage{booktabs, amssymb}
\usepackage{tabularx}
\usepackage{multirow}
\usepackage[table]{xcolor}
\usepackage{mathptmx}                  
\usepackage{booktabs}
\usepackage{longtable}
\usepackage{array}
\usepackage{amssymb}
\usepackage{amsmath}

\newcolumntype{L}[1]{>{\raggedright\arraybackslash}p{#1}}
\newcolumntype{C}[1]{>{\centering\arraybackslash}p{#1}}

\onlineid{0}

\vgtccategory{Research}

\vgtcinsertpkg

\title{User Preferences for UI Anchoring in MR: Effects of Task Mobility and Interface Properties}

\author{Jo\~ao Belo\thanks{e-mail: joaobelo92@gmail.com}\\ %
     \parbox{1.6in}{\scriptsize \centering Saarland Informatics Campus \\ Saarland University}%
\and Sina Elahimanesh\thanks{e-mail: siel00002@uni-saarland.de}\\ %
     \parbox{1.6in}{\scriptsize \centering Saarland Informatics Campus \\ Saarland University} %
\and Anna Feit\thanks{e-mail: feit@cs.uni-saarland.de}\\ %
     \parbox{1.6in}{\scriptsize \centering Saarland Informatics Campus \\ Saarland University}} %

\teaser{
  \centering
  \includegraphics[width=\linewidth]{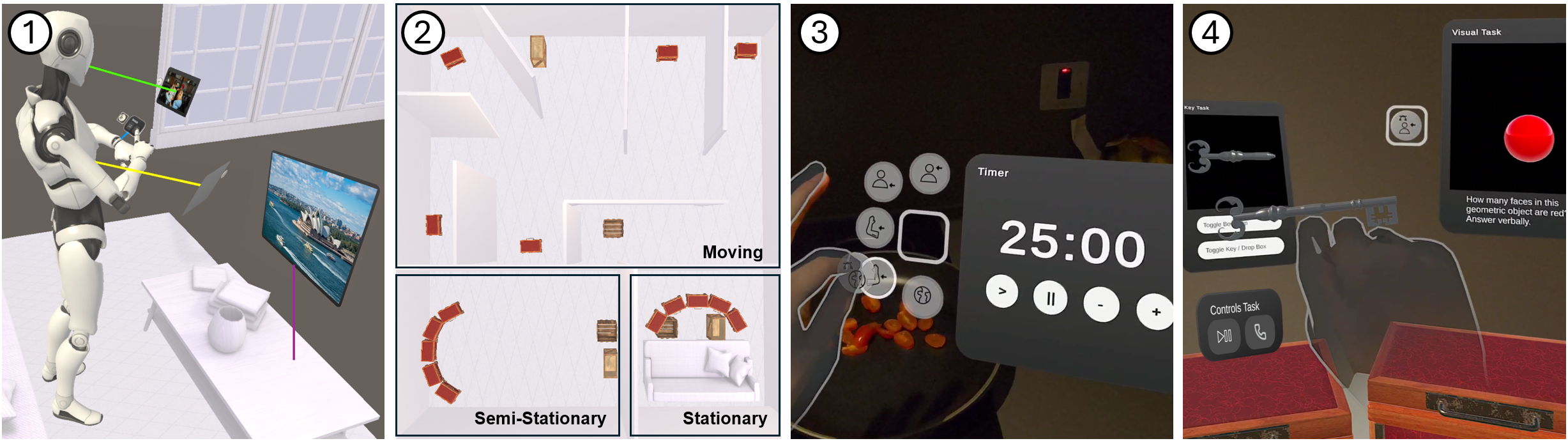}
  \caption{Overview of our MR study investigating user preferences for UI anchoring across differing mobility settings and interface properties. (1) Anchoring reference frames explored in the study: head (green), torso (yellow), arm (blue), and world (purple). (2) Mobility conditions in the experiment : moving, semi-stationary, and stationary. (3) Participants freely selected anchoring modes during the study switching between reference frames at any time. (4) The three interface types used in the experiment.}
  \label{fig:teaser}
}

\abstract{
Anchoring - the choice of frame of reference for mixed reality (MR) interface elements - is a critical design decision involving trade-offs between accessibility, interaction comfort, and visual interference.
Despite its importance, user preferences for anchoring across different mobility contexts and interface properties remain poorly understood, as prior work has largely focused on specific tasks or fixed interface configurations. 
We address this through an exploratory mixed-methods user study in which participants configure anchoring strategies across different mobility conditions and interface types.
Combining behavioral analysis with structured qualitative inquiry, we analyze how participants select and reason about anchoring modes.
Our results show a clear transition from world-anchored interfaces in stationary contexts to body-anchored interfaces during locomotion. However, no single body anchor consistently dominates, highlighting the personal nature of anchoring strategies.
Our qualitative analysis reveals the factors users consider in their  anchoring decision, including interface accessibility, stability during interaction, visual clutter, and individual mental models. These findings inform the design of adaptive and controllable MR interfaces and highlight the importance of supporting user customization. 
} 

\keywords{Mixed Reality, 3D user interfaces, Interaction.}

\begin{document}


\firstsection{Introduction}

\maketitle

Ongoing progress in Mixed Reality (MR) hardware is steadily advancing the vision of ubiquitous, everyday MR \cite{abrash2021future}.
As these systems move beyond controlled settings into daily use, the usability of MR user interfaces (UIs) becomes a priority.
A fundamental design decision in this context is the UI frame of reference - that is, how virtual content is \emph{anchored} relative to the user, their body, and the surrounding environment (\autoref{fig:teaser}.1).

Different \emph{anchoring modes} offer distinct benefits and drawbacks that can substantially affect usability.
World-anchored UIs are spatially stable and support spatial understanding, but may become difficult to access or stay outside the user's field of view as they move or change their posture.
In contrast, body-anchored UIs ensure constant availability and visibility, but can introduce visual clutter and interaction challenges as user movements cause the interaction target to shift during input.
As an example, consider a messaging interface that is anchored to the environment.
While such anchoring may work well when the user is sitting at their desk, the interface quickly becomes inaccessible if the user stands up (e.g. when using a standing desk) or starts walking (e.g. because they are running late to a meeting) while trying to read an incoming message.
Conversely, anchoring the messaging interface to the user's head keeps it in constant view which may be convenient during locomotion, but can quickly become intrusive when the user stops to talk to a colleague and UI elements clutter their field of view.
In both cases, inadequate anchoring of MR UI elements can disrupt a task in the real-world or  impair the user experience with MR interfaces - particularly in situations where users frequently transition between sitting, standing, and walking.

In this work, we explore how mobility and interaction modality shape end users' preferences for UI anchoring in MR.
Previous work has mainly focused on quantifying users' task performance with different anchoring modes in isolated task conditions.
Across studies, results vary and may even appear contradictory, likely due to differences in experimental choices such as mobility conditions, interaction modalities, and task characteristics \cite{manakhov2024gaze, rasch2025track}.
These observations suggest that such factors play a critical role in determining appropriate anchoring strategies.
Rather than focusing on performance metrics, our goal is to develop a principled understanding of \emph{anchoring preferences} across different mobility and interaction conditions.
Therefore, we reviewed the MR literature and identified world-, head-, torso-, and arm-anchoring as the most common and promising anchoring modes.
We designed and implemented intuitive anchoring customization controls that allow users to change and iteratively refine the anchoring mode of MR interfaces during runtime, rather than relying on fixed, designer-defined choices (\autoref{fig:teaser}.3).
This enables users to directly experience and control the trade-offs of different anchoring strategies as their context and activity change.
Finally, we designed representative interaction scenarios that span different mobility conditions and interactive tasks to explore anchoring preferences across a range of everyday MR use cases.
Specifically, we selected three UI elements reflecting distinct interaction modalities: (1) an instruction panel combining visual guidance with hand-based input, (2) a view panel supporting purely visual tasks, and (3) a control panel designed primarily for manual input.

Notably, our work investigates anchoring in interfaces that rely exclusively on hand input.
Hand input is particularly compelling for everyday MR scenarios, as it eliminates the need for dedicated controllers and is widely supported in modern devices (e.g., Apple Vision Pro, Meta headsets).
UI anchoring plays a more critical role in maintaining interaction stability and usability under movement, as hand interaction is coupled to the user's hand and body motion, unlike controller input which relies on a stable intermediary.

To investigate the effects of mobility on anchoring preferences, we compared usage patterns and subjective user preferences across three distinct mobility settings in MR: 1) stationary, where the user sits in a chair; 2) semi-stationary, where the user performs tasks across two distinct zones in the room; and 3) moving, where the user performs tasks across various points in the room (\autoref{fig:teaser}.2).
Participants were required to complete different tasks using the UIs introduced above and were given complete control to customize the interface anchoring according to their preference throughout the study (\autoref{fig:teaser}.3 and \ref{fig:teaser}.4).

The study revealed that when users are stationary, world-anchoring is unequivocally preferred across all task modalities we tested.
However, as the level of mobility increased, there was a shift in preference towards body-anchoring, with the semi-stationary setting showing a larger variety of preferences between world- and body-anchoring options.
Additionally, our analysis revealed that anchoring preferences are formed early and can be widely different across individuals.
We also contribute a thematic analysis of semi-structured interviews, where participants reasoned about their anchoring decisions and noted that aspects such as stability in relation to the environment, visual clutter, and UI properties such as the interaction method and size influence their anchoring decisions.
These results underscore an important finding - optimal anchoring is not universal.
Instead, it depends on the level of motion, the nature of the task, interaction modality, UI properties, and individual user preferences.
This highlights the critical need for anchoring mechanisms that are not only adaptive but also user-customizable to ensure the usability of future MR systems.
This work contributes to understanding how users prefer to anchor MR interfaces in everyday contexts and the implications of different types of UIs.

\section{Related Work}

\subsection{Spatial Reference Frames for UI Anchoring}
The design of three-dimensional user interfaces (3D UIs) in Mixed Reality has a direct impact on usability and user experience.
A major design decision in this space concerns the placement of virtual content and how the UI position changes over time.
The first explorations of different anchoring modes date back to 1993, when Feiner et al. created an MR system capable of anchoring UIs to the headset, the user, or objects and locations in the physical world \cite{feiner1993windows}.
Since then, researchers continued to investigate UI anchoring in MR, contributing different perspectives on the topic. 
Foundational discussions of egocentric and exocentric reference frames in 3D interaction appear in general 3D UI literature \cite{LaViola3d}, which conceptualize how users perceive and navigate 3D spaces. 
Building on these spatial reference concepts, Ens et al. proposed a taxonomy where UIs are classified as egocentric or exocentric, explicitly introducing the capability to move a UI relative to its frame of reference \cite{ens2014ethereal}.

More recent works explored specific settings, where it is common to anchor the content to the environment \cite{weizhou2024strategies}, or to task-specific tools or objects (e.g., surgery) \cite{acherki2025arsurgery}, or to the body, such as when walking \cite{rasch2025track} or to resemble  smartphone-style interaction \cite{zhu2024phonevr} through wrist-anchoring.
Past research shows that people navigate space and interpret reference frames based on their point of view \cite{torok2014reference}.
This influences how users form mental maps and interact with virtual elements over time.
As a result, the effectiveness of a given anchoring strategy depends on the task being performed, the interaction modality, and the extent of user movement.
This dependence may explain the mixed and sometimes conflicting findings reported in earlier studies.
In our work, we focus on world- and user-body anchoring, providing an overview of both.

\textit{World}-anchored interfaces remain stable within the environment and define reference frames relative to the physical world.
Previous work shows that such anchoring often supports faster task completion and reduces cognitive load \cite{ghasemi2021headlocked, tabone2024insights}.
For instance, Manakhov et al. \cite{manakhov2024gaze} demonstrated that target acquisition is faster and precision improves as the target's stability relative to the physical environment increases.

In contrast, \textit{head}-, \textit{body}-, \textit{torso}-, and \textit{arm}-anchored interfaces define egocentric reference frames that move with the user.
Head-anchored interfaces, which remain fixed in the user’s field of view, are well suited for quick-glance elements such as notifications or menus, supporting fast access but potentially increasing interference during demanding dual-task scenarios, as shown for walking pedestrians \cite{rasch2025track, tabone2024insights} and for cognitive load from anchored text \cite{mack2023headplacements}.
Body- and torso-anchored interfaces offer a more stable yet still accessible alternative, allowing repeated interaction without constant head movement, though sometimes at the cost of reduced walking efficiency or interaction accuracy \cite{jannat2025handUIs, rasch2025track, casadio2012bodymachine}.
Arm- or hand-anchored interfaces follow the user’s limb and support natural pointing and manipulation, which can improve comfort and precision during repeated interactions, while also introducing additional cognitive demands during pointing \cite{li2021armstrong} and combined cognitive-motor load while walking \cite{jannat2025handUIs}.

\textit{Object}-anchored interfaces bind virtual elements to physical objects in the environment, enhancing contextual awareness \cite{xr_objects, praschl2022geo, pei2024mobility}.
By aligning digital information with relevant real-world items, object anchoring can improve understanding and task performance \cite{acherki2025arsurgery, wand2025ARaquisition, rometsch2022exploration}.
Prior studies show that object-based anchoring increases engagement and precision, particularly in professional and safety-critical scenarios \cite{acherki2025arsurgery, park2020augmented}.

Due to the trade-offs between different anchoring modes, some works have proposed adaptive approaches that adjust anchoring based on the state of the user or context \cite{davari2024context, lindlbauer2019context}. 
However, these methods mostly rely on system-driven rules and do not consider what users prefer.
Little is known about user preferences during continuous movement or across different UI types and tasks, which we investigate in our work.

\subsection{User-Driven UI Anchoring and Positioning}
The ability for users to control the position and anchoring of UIs in MR systems is crucial as it directly affects usability, cognitive load, and physical safety \cite{Shneiderman2018DesigningUI, LaViola3d}.
Existing work has proposed interaction techniques to reposition UI elements that enable users to explicitly manipulate their spatial coordinates \cite{hayatpur2019spatial}.
Beyond manual control, automatically adjusting the UI position at runtime within a specific frame of reference has also been explored using constraint-driven \cite{bell2001view} methods and optimization approaches \cite{lindlbauer2019context, belo2022auit}.
Lu and Yan explored transition mechanisms that combine different degrees of automation and user manipulation \cite{lu2022transitions}.
Pei et al. \cite{pei2024mobility} propose a taxonomy distinguishing interfaces attached to static, dynamic, or self-referenced entities, and their FingerSwitches technique shows that manual transitions between anchor states can improve usability, although effectiveness depends on the surrounding environment.
Cross-reality and asynchronous task systems further demonstrate the value of maintaining continuity across spatial and temporal contexts \cite{cho2025bridging}, while mobile AR frameworks emphasize smooth repositioning to support natural interaction in dynamic settings \cite{cao2023mobileAR}.
These works show the importance of mobility, context, and adaptable anchoring for interaction comfort and usability.
Despite these advances, most studies focus on automatic or fixed anchor transitions without examining how users themselves customize anchoring preferences across varying mobility levels, interaction modalities, and UI types. In our study participants can control how interfaces update their position over time and explores users' customization behavior and their reasoning behind anchoring choices.

Recent work highlights the importance of adaptability in MR interfaces \cite{lindlbauer2022future}.
Systems that incorporate user feedback to personalize layouts show measurable benefits, even when user goals are uncertain or only partially defined \cite{johns2023adaptations, johns2023towards}.
VR and MR guidelines similarly emphasize the need to accommodate ergonomic differences \cite{belo2021xrgonomics}, task diversity \cite{lindlbauer2019context}, and varied use contexts, advocating for customizable interfaces rather than fixed layouts.
Researchers have used optimization to adapt MR interfaces leveraging virtual-physical semantic relationships \cite{cheng2021semanticadapt}, physical surroundings \cite{cheng2023interactionadapt}, and social and environmental context \cite{zhipeng2024situationadapt}.
MineXR elicits personalized MR configurations directly from users, revealing substatial variation even for comparable tasks \cite{cho2024minexr}.
Studies of MR-based customization within retail environments also show flexibility is valued by users and industry stakeholders \cite{jin2025adoptingMR}.
More broadly, Krauß et al. \cite{krauss2021MRdesign} synthesize community perspectives on MR design recommendations, emphasizing that spatial configuration and reference frame choices are central open design questions.
Domain-specific applied work echoes these findings, with Jusko et al. \cite{jusko2025holographic} showing how holographic cue placement affects usability in aviation tasks.
FlowAR \cite{jo2023flowar} adapts AR-overlay placement to support full-body movement during yoga exercises without disrupting exercise flow, illustrating how interface placement can respond to movement demands.

However, most prior studies employ fixed anchoring schemes.
In contrast, our work emphasizes user customization of UIs with different anchoring choices.

\section{Study}
We designed an exploratory mixed-methods study to understand how users' preferences for anchoring of UI elements changes across different usage scenarios and settings.
Concretely, we aim to answer the following research questions:
\begin{enumerate}
    \item[RQ 1] How does the user's mobility setting (e.g., stationary, moving) influence their anchoring preferences during MR tasks?
    \item[RQ 2] How do different interface types (e.g., visual display, control panel) affect the user's anchoring preferences? 
    \item[RQ 3] What factors do users weigh when selecting an anchoring strategy, and how do these factors interact with mobility context and interface type?
\end{enumerate}

\begin{figure}[t]
  \centering
  \includegraphics[width=0.9\linewidth]{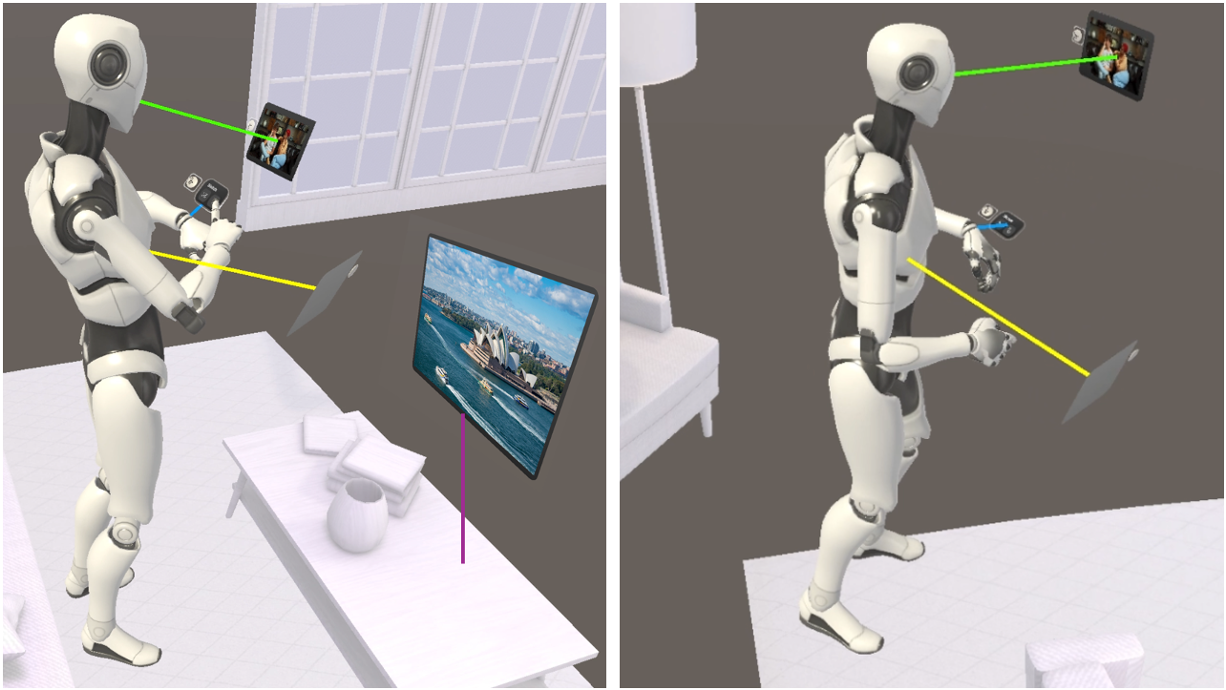}
  \caption{Behavior of the anchoring modes explored in our system as the user’s pose changes over time. Interfaces may be anchored to the head (green), torso (yellow), arm (blue), or the world (purple).}
  \label{fig:anchors}
\end{figure}

\subsection{Anchoring Modes}
The anchoring targets in our work (also referred to as spatial reference frames in related work) were chosen based on successful use in prior MR systems and literature, feasibility within existing consumer hardware, and insights gathered through iterative prototype design. Specifically, we consider the following four anchoring modes, illustrated in Figure~\ref{fig:anchors}.

\textbf{World}  anchoring ties the UI's position to the physical environment using real-world feature points identified by the headset's tracking system.  As the headset continuously recalculates its pose relative to these feature points, virtual objects remain spatially stable in the environment. 
    This is the standard anchoring mode for persistent UIs in MR operating systems such as the VisionOS and the Meta Horizon OS.
    
\textbf{Head}  anchoring attaches the UI element to the headset’s local coordinate system, maintaining a fixed position relative to the user’s field of view (FoV). 
    While this ensures high visibility, it may contribute to visual clutter when the UI is not in active use. 
    In pilot studies we also experienced an effect of head jitter on target acquisition, as subtle head movements resulted in minor UI jitter which often disrupted input stability during interaction.

\textbf{Torso}  anchoring places the UI within the torso's local coordinate system. 
    We use the body tracking capabilities of the Movement SDK to estimate the torso's coordinate system. 
    Torso anchoring reduces persistent visual clutter, as the UI remains out of the FoV unless the user intentionally turns toward it.
    While modern headsets can reliably perform body tracking, our pilot studies showed that certain UI placements induced unintended torso movement during hand interaction that could also result in UI jittering disruptions.

\textbf{Hand}  anchoring locks UI elements to the user's left or right hand, using the selected wrist’s local coordinate system as the spatial reference. 
    This enables users to summon interfaces quickly and mitigates stability issues observed with head and torso anchoring. 
    However, it requires bimanual coordination for interaction: one hand positions the UI while the other interacts with it. 
    Additionally, the effective space available for interface placement is smaller compared to the other anchoring options.

\begin{figure*}[t]
  \centering
  \includegraphics[width=0.75\linewidth]{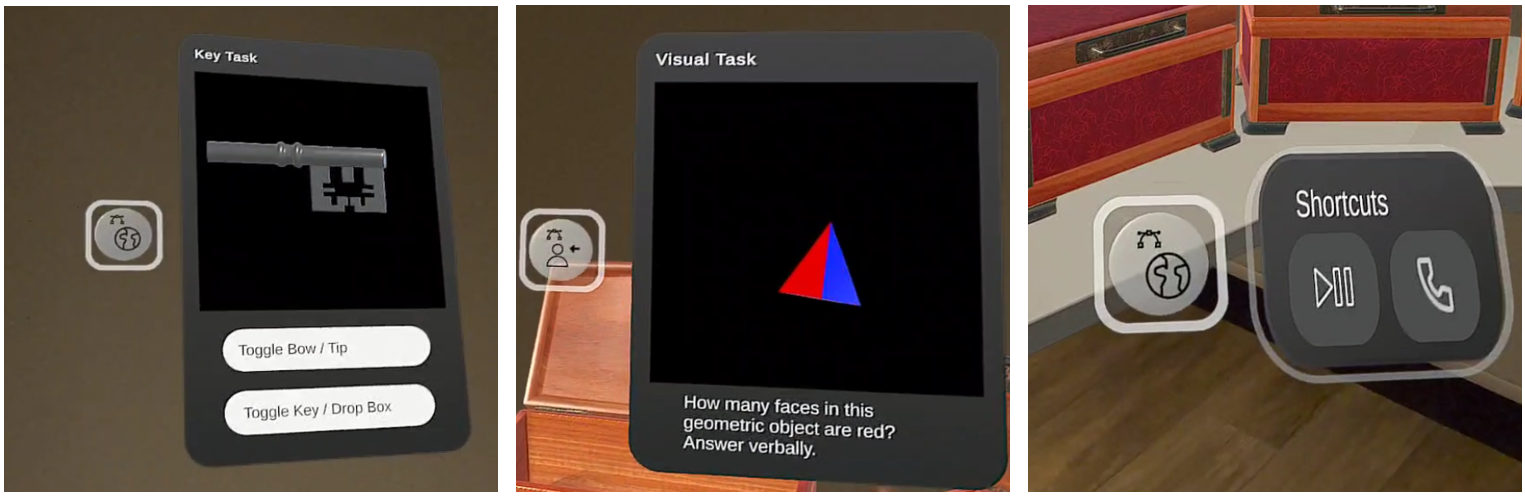}
  \caption{Interfaces used in the study. \textbf{Left:} Key task. \textbf{Center:} Visual task. \textbf{Right:} Controls.}
  \label{fig:interfaces}
\end{figure*}

During design and prototyping, we explored variants similar to HeadDelay \cite{manakhov2024gaze} and other hybrid approaches.
Pilot studies suggested that introducing movement-dependent delays or dynamically toggling between head and world (for head anchoring) or torso and world (for torso anchoring) based on customizable motion thresholds might improve stability during hand interaction.
However, these behaviors were sensitive to implementation details, and the design space grew combinatorially with each added parameter.
To avoid an unmanageable number of experimental conditions, we opted for the simplest implementation of each anchoring mode.
This choice ensures a clean comparison in which the inherent characteristics of each frame of reference can be examined more directly.

To enable quick customization of UI anchoring, we implemented a button that presents all anchoring options and allows users to switch modes on demand (see \autoref{fig:teaser}.3). 
In contrast to existing work, our prototype allows the user to adjust the position of the UI element within the chosen reference frame - i.e., the local coordinates relative to the specific anchor point.
Providing this level of control reduces the friction associated with fixed or designer-determined UI placements and accommodates individual differences in preference, ergonomics, and situational context.
Furthermore, user-controlled placement ensures that comparisons across anchoring modes remain focused on the anchoring behavior itself rather than confounded by discomfort or suboptimal layouts.


\subsection{Study design}

Our exploratory mixed-methods study uses a within-participants design with three conditions corresponding to the participants' \emph{mobility} (stationary / semi-stationary / moving). We use a repeated-measures design, where participants go through multiple task trials within each mobility condition. In each trial, they are shown three different \emph{interface types} (key / visual / controls) each corresponding to a different experimental task which participants experience throughout each trial. In the following, we describe the mobility condition and interface types in detail

\subsubsection{Mobility}
\label{sec:mobility}
In our work, the mobility setting captures the degree to which users are physically moving while interacting with MR interfaces.
This factor is particularly important in MR, as motion can influence perceptual stability and interaction performance, which in turn affects how users prefer different anchoring options. 
Everyday MR scenarios often envision replacing traditional smartphone apps in \textbf{moving} contexts - for example navigation, or messaging while walking at a comfortable pace \cite{abrash2021future}.
At the same time, many compelling use cases involve remaining \textbf{stationary}, such as enhancing productivity at a workstation with virtual monitors \cite{cheng2025augmented} or standing at a desk using MR to sketch, sculpt, or arrange virtual design elements. 
Consider a \textit{cooking scenario}, where users naturally move between different areas of the kitchen.
Virtual UI elements can assist across multiple zones: checking an ingredient list by the fridge, watching a video tutorial by the counter, or starting a timer by the stove (see video accompanying the paper). 
Motivated by this, we also explore a \textbf{semi-stationary} condition, in which users perform tasks by going back and forth between two zones.
\autoref{fig:teaser}.2 illustrates each mobility condition.

\subsubsection{Interface Types}
We designed three experimental tasks that aim to approximate realistic MR interactions while maintaining experimental control across the different mobility settings.
Each task operationalizes a distinct interface type that reflects common MR use cases. Users experience tasks in parallel, capturing typical multitasking demands in MR environments, where users must coordinate virtual interactions while simultaneously attending to the physical world.
For each task, we designed and implemented an MR user interface with different interaction properties. The three interface types are shown in \autoref{fig:interfaces} and explained in the following. 
They were designed to differ in input modalities, interaction frequency, urgency, size, and whether the task is location-bound, mimicking realistic MR interface demands and maximizing the chance to observe differences in anchoring preferences. However, this design  does not allow us to attribute observed differences to any single interface property and we only treat interface type as the experimental factor in this study. The tasks corresponding to each interface type were selected to reflect emerging MR form factors, where headsets resemble everyday glasses and interaction occurs without handheld controllers.

\textbf{Key.}  
The key task (\autoref{fig:interfaces}, left) requires both visual attention and hand input, has low urgency but a high interaction frequency since it it is done throughout each trial. It has a large interface size and is location-bound. 
Motivated by interaction patterns common in Meta Horizon OS, this task combines visual guidance with direct hand input, reflecting structured, goal-oriented real-world workflows such as guided cooking or product assembly.
In a search task, participants must identify a target key among several alternatives placed inside closed boxes and place it in a target box.
The interface displays either the correct key handle, the correct key tip, or the target box. The participant must toggle between them by pressing the corresponding buttons (see \autoref{fig:interfaces}, left). 
The spatial arrangement of the boxes determines the mobility demands of the task, requiring participants to move within the environment to inspect and retrieve objects (see section \ref{sec:mobility} below).
For better experimental control, we simulate the real-world elements (boxes, keys, drop box) as well as distractor walls in the moving condition (see \autoref{fig:teaser}.2).

\textbf{Visual.}  
The visual task only requires visual attention and no hand input. It has a low interaction frequency and is not location bound. It represents common background activities such as glancing at a news ticker or periodically checking informational content such as a notification center (\autoref{fig:interfaces}, middle).
The interface displays a rotating geometric object with colored faces (sphere, four-sided pyramid, five-sided pyramid, cube, or octahedron).
At random intervals, participants are prompted to verbally report the number of faces colored red.

\textbf{Controls.}  
The controls interface requires manual input only. It has a low interaction frequency and small interface size and is not location bound. However, it has a high urgency. 
The controls interface emulates intermittent system-level interactions that users perform throughout the day, such as answering calls, controlling media playback, or adjusting notification settings (\autoref{fig:interfaces}, right).
During the experiment, participants receive simulated events (e.g., call that must be answered immediately or music playing that should be stopped) and must use the buttons through hand poking to handle them.

\subsection{Apparatus}
We implemented our prototype in Unity 6000.2, with all user interfaces developed using the Meta Core and the Meta Interaction SDK. 
All interactions relied on hand tracking and poke-based input, enabling users to directly engage with interfaces through direct touch using their hands.
To ensure a consistent and cohesive interaction experience aligned with application patterns on the Meta Quest we used the Meta UI Set to develop the UIs.

The prototype operated in Mixed Reality using passthrough, providing real-time visualization of the surrounding physical world.
Additionally, the Movement SDK was used to capture participant movement data and to access the Torso coordinate system, which supported torso anchoring.

All sessions were run using the Meta Quest 3 headset over Air Link on a high-end computer, to provide participants a smooth untethered experience.
The experiment took place in a quiet, well-lit room, with a chair positioned at the center to serve as the starting point for each condition.

\subsection{Participants}
We recruited 19 participants 
(10 male, 9 female) ranging  from 23 to 60 (mean: 36.52; std. dev: 10.97) in age.
They all reported normal or corrected to normal vision and no walking impairments. 
All participants provided written informed consent before starting the study and received a monetary compensation for their participation.
Participants reported relatively low prior experience with MR devices (M=1.53, SD=0.84, on a 1-7 scale). 

\subsection{Procedure}
Each experimental session lasted approximately 90 minutes.
Upon arrival, participants were welcomed and given an overview of the study objectives, experimental tasks, and types of data we collected.
After signing a consent form, participants completed a demographics questionnaire, including a self-report measure of prior MR experience.

The session began with a training phase in which participants were familiarized with the MR interface and how to reposition interface elements and switch between anchoring modes.

Participants then completed three experimental blocks corresponding to the three mobility conditions: \textit{stationary}, \textit{semi-stationary}, and \textit{moving}.
The order of the three mobility conditions was counterbalanced using all six possible orderings.
Five orderings were assigned to three participants each, and one ordering was assigned to four participants.
Each block contained 10 trials, each beginning with a new key task.
During the trial, participants were also prompted to perform with one visual and one controls task via an audio cue occurring at a random time between 15 and 30 seconds after trial started. 
Because every trial included all three interface types - the Key task ran continuously while the Visual and Controls tasks were triggered by audio cues at random intervals - the interface types did not require separate counterbalancing.

Each block consisted of ten trials.
The first three served as exploration and familiarization trials, during which participants were explicitly instructed and encouraged to try all four anchoring methods.
After each practice trial, all interfaces were reset to a default configuration (all set to world anchoring) to ensure equal exposure to all anchoring options.
Before proceeding, the experimenter verified that participants had tried each anchoring mode at least once, ensuring that all participants experienced all the anchoring modes.

The remaining seven trials were treated as experimental trials and were completed consecutively.
Before starting these trials, participants were informed that their behavior would be recorded and were instructed to configure the interfaces using their preferred anchoring modes.
Participants were reminded that they could change anchoring configurations at any point during the trials.
Only data from these seven trials were included in the analysis.

After completing each mobility condition, participants took a short break and completed the questionnaire and the semi-structured interview.

At the conclusion of the session, participants were debriefed about the study’s goals, thanked for their participation, and compensated with €20.
The study was conducted in the participant's native language (Portuguese) and the questionnaire material was translated by a native speaker from English.
Before analysis, all answers to open-ended questions were translated back to English.

\subsection{Data collection and Analysis}
\label{sec:data-analysis}

After each mobility condition, participants completed a structured questionnaire and participated in short semi-structured interviews in which they articulated the reasoning behind their anchoring choices.
The questionnaire asked participants to rate their perceived ease-of-use of each anchoring method for each interface type on a 7-point likert-scale with questions adapted from UMUX~\cite{finstad2010umux}. Additionally, it asked participants to rank the anchoring method according to their preferences for each interface type in the specific mobility condition.

To establish significant differences between conditions, we used non-parametric repeated-measures analyses to match the ordinal nature of the ratings, the within-participants study design, and the repeated ratings across interface types and anchoring modes.
Specifically, to compare the four anchoring methods (world, head, torso, and arm) \emph{within} each mobility-condition $\times$ interface-type combination, we used Friedman tests, which are appropriate for comparing more than two related samples of ordinal data, and report Kendall's $W$ as the omnibus effect-size measure.
Significant omnibus tests were followed by paired Wilcoxon signed-rank comparisons with Holm correction. Rank-biserial correlation, reported as $r_{rb}$, was used as the pairwise effect-size measure.
We used the same approach to compare the three mobility conditions (stationary, semi-stationary, and moving) within each anchoring-method $\times$ interface-type combination. 
The significance threshold was $\alpha=.05$. Consistent with the non-parametric repeated-measures tests, we report ease-of-use ratings using medians and interquartile ranges (IQRs). 
Full omnibus and post-hoc results, including significant and non-significant comparisons, Holm-corrected $p$-values, and effect sizes, are provided in~\autoref{app:full-statistics}. In the results we only present the most interesting effects.

Following the structured questionnaire, we conducted a semi-structured interview to where we asked participants to explain their anchor choices for each interface type and their overall impressions of the different anchoring options.
See ~\autoref{app:details} for the full set of questions. Full responses are also included in the supplementary material.
We employed a reflexive thematic analysis \cite{braun2006thematic, braun2019reflexive} to analyze participants' answers. 
Two researchers (authors of this paper) analyzed 171 explanations (19 participants $times$ 3 mobility conditions $times$ 3 interface types )
using an inductive, bottom-up coding approach.
The two researchers independently performed an initial coding of the responses and iteratively compared and refined the emerging categories through discussion.
Through this process, the coders consolidated overlapping codes and agreed on a set of themes capturing the main reasoning patterns underlying participants’ anchoring decisions.

In addition, we recorded task completion times and accuracy metrics for each interface.
However, participants were explicitly instructed to prioritize comfort and correctness over speed. 

\section{Results}
In this section, we report the results of our user study. 
We begin with a quantitative analysis of the questionnaire responses and 8778 task completion events logged throughout the trials.
As the completion time and accuracy metrics did not yield conclusive results, we do not discuss them in detail.
Statistical procedures are detailed in Section \ref{sec:data-analysis}.

\subsection{Anchoring Preferences}
Our primary goal is to understand anchoring preferences under different locomotion conditions and how these preferences vary across interface types. 
Therefore, we begin by analysing questionnaire data capturing participants' declared favourite anchor choices by mobility condition and interface type.
Table~\ref{tab:preference-distribution} shows users' declared anchor preferences, revealing mobility and task-dependent patterns.
Overall, the results indicate that locomotion primarily determines the category of preferred anchoring (world- vs. body- anchoring), whereas interface type drives the choice of specific body anchor during movement.

\begin{table}[h]
  \centering
  \caption{Distribution of top anchor choices (percentage of participants) by condition and task.}
  \label{tab:preference-distribution}
  \small
  \setlength{\tabcolsep}{4pt}
  \renewcommand{\arraystretch}{1.15}
  \begin{tabularx}{\columnwidth}{
    >{\centering\arraybackslash}X
    >{\centering\arraybackslash}X
    >{\centering\arraybackslash}X
    >{\centering\arraybackslash}X
    >{\centering\arraybackslash}X
    >{\centering\arraybackslash}X
  }
    \toprule
    \textbf{Mobility} & \textbf{Interface} & \textbf{World} & \textbf{Head} & \textbf{Torso} & \textbf{Arm} \\
    \midrule
    \multirow{3}{*}{\textbf{Stationary}}
      & Key      & \cellcolor{green!100}\textbf{100.0} & \cellcolor{green!0}0.0  & \cellcolor{green!0}0.0  & \cellcolor{green!0}0.0 \\
      & Visual   & \cellcolor{green!100}\textbf{100.0} & \cellcolor{green!0}0.0  & \cellcolor{green!0}0.0  & \cellcolor{green!0}0.0 \\
      & Controls & \cellcolor{green!95}\textbf{94.7}   & \cellcolor{green!0}0.0  & \cellcolor{green!0}0.0  & \cellcolor{green!5}5.3 \\
    \midrule
    \multirow{3}{*}{\textbf{Semi-Stat.}}
      & Key      & \cellcolor{green!42}\textbf{42.1}   & \cellcolor{green!32}31.6 & \cellcolor{green!16}15.8 & \cellcolor{green!11}10.5 \\
      & Visual   & \cellcolor{green!5}5.3              & \cellcolor{green!42}42.1 & \cellcolor{green!47}\textbf{47.4} & \cellcolor{green!5}5.3 \\
      & Controls & \cellcolor{green!5}5.3              & \cellcolor{green!21}21.1 & \cellcolor{green!16}15.8 & \cellcolor{green!58}\textbf{57.9} \\
    \midrule
    \multirow{3}{*}{\textbf{Moving}}
      & Key      & \cellcolor{green!5}5.3              & \cellcolor{green!42}\textbf{42.1} & \cellcolor{green!37}36.8 & \cellcolor{green!16}15.8 \\
      & Visual   & \cellcolor{green!0}0.0              & \cellcolor{green!32}31.6 & \cellcolor{green!47}\textbf{47.4} & \cellcolor{green!21}21.1 \\
      & Controls & \cellcolor{green!0}0.0              & \cellcolor{green!26}26.3 & \cellcolor{green!16}15.8 & \cellcolor{green!58}\textbf{57.9} \\
    \bottomrule
  \end{tabularx}
\end{table}

When stationary, participants overwhelmingly preferred world anchoring across all interface types, accounting for 94.7–100\% of preferred anchor selections.
As locomotion is introduced, preferences shift progressively from world to body anchoring.
For the Visual and Controls task, this shift occurs as soon as locomotion is involved.
In contrast, we still observe that 42.1\% of the participants prefer world anchoring for the Key task in semi-stationary settings, even though the task must be completed across two locations.
This suggests that the spatial relationship between the task’s visual instructions and the corresponding real-world elements (simulated in our study) remained a dominant factor in participants’ anchoring choices in the semi-stationary condition.

In the moving condition, preferences shift entirely toward body anchoring across all tasks.
However, no single body anchor (Head, Torso, or Arm) universally dominates across tasks.
For the Key and Visual tasks, preferences are distributed across different body anchors, suggesting that multiple placements can support these interactions.
In contrast, the Controls task shows a different pattern.
Arm anchoring was chosen by 57.9\% of participants as their preferred anchoring mode, substantially higher than for Key (15.8\%) or Visual (21.1\%) tasks.
This preference reflects the task’s requirement for occasional, rapid manual input: keeping the interface close to the hand minimizes reach time and interaction latency.

\begin{figure*}[t]
  \centering
  \includegraphics[width=\linewidth]{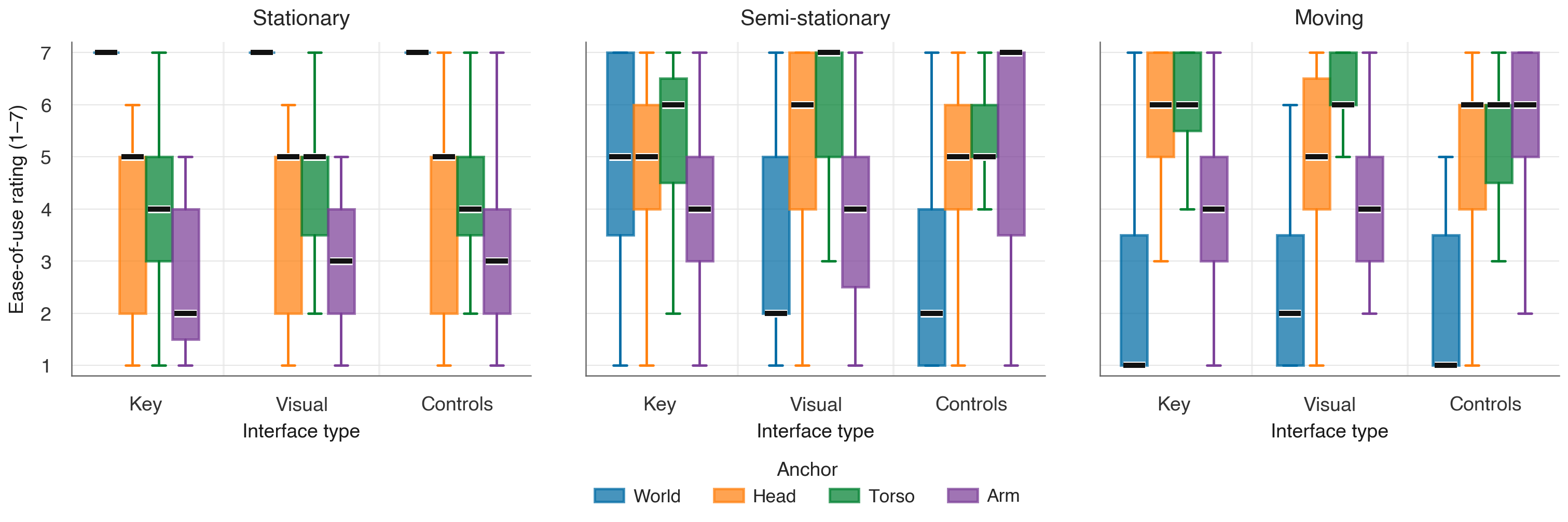}
  \caption{Distribution of ease-of-use ratings for the four anchoring methods across interface types, shown separately for the Stationary, Semi-Stationary, and Moving conditions. Black horizontal ticks mark the median of each distribution. Higher values indicate greater perceived ease of use. Note that for the stationary condition, world-anchoring was rated uniformly high (7) for all interface types.} 
  \label{fig:rating-distributions}
\end{figure*}

To understand anchoring preferences further, we also asked participants usability questions from UMUX on a Likert scale from 1 to 7 \cite{finstad2010umux}.
The ease-of-use ratings provide complementary evidence for above mentioned preference patterns.
Figure~\ref{fig:rating-distributions} shows the distribution of ease-of-use ratings across the nine mobility condition $\times$ interface type combinations.
It reveals a clear gradient where usability ratings of world-anchor dominate strongly in the stationary condition (\autoref{fig:rating-distributions}, left), through mixed ratings across world- and body-anchors in the Semi-Stationary condition (\autoref{fig:rating-distributions}, middle) to body-anchors being rated most highly in the Moving condition (\autoref{fig:rating-distributions}, right).
This progression highlights that mobility level is the primary determinant of the perceived usability of anchoring strategy between world and body anchoring modes, and 
Friedman tests comparing world-anchor ease ratings across mobility conditions, paired within participants, showed significant differences for all three interface types ($\chi^2 = 28.56$-$32.38$, all $p < 10^{-6}$, Kendall's $W = 0.75$-$0.85$).

In the Stationary condition, Head and Torso were generally rated positively, with medians between 4.0 and 5.0, while Arm anchoring was the main exception, with medians between 2.0 and 3.0.
When transitioning from Stationary to Moving, usability ratings showed a pronounced shift: World anchoring dropped to near-unusable levels, with medians between 1.0 and 2.0 and IQRs spanning 1.0-3.5, whereas body-anchored references received substantially higher median ratings between 4.0 and 6.0.
Complete medians and IQRs are reported in Table~\ref{tab:ease-descriptives} in \autoref{app:full-statistics}.
Friedman tests within each mobility condition and task (Stationary/Semi-Stationary/Moving $times$ Key/Visual/Controls) indicated that anchor choice significantly affected perceived ease in all nine combinations (all $p < .05$, $n = 19$). Effects were largest when participants were stationary (Key: $\chi^2 = 40.72$, $p < 10^{-8}$, $W = 0.71$; Visual: $\chi^2 = 40.42$, $p < 10^{-8}$, $W = 0.71$; Controls: $\chi^2 = 31.52$, $p < 10^{-6}$, $W = 0.55$).


The Semi-Stationary condition exhibited a transitional pattern bridging the extremes.
Perceived usability ratings were distributed across both World and body-anchored options, with Torso anchoring emerging as particularly versatile.
Torso received a median rating of 7.0 ($IQR = 5.0$-$7.0$) for the Visual interface, 6.0 ($IQR = 4.5$-$6.5$) for Key, and 5.0 ($IQR = 5.0$-$6.0$) for Controls.
For Controls, Arm received a higher median of 7.0, but ratings were also more widely distributed ($IQR = 3.5$-$7.0$).
Holm-corrected Wilcoxon signed-rank post-hoc tests confirmed that Torso was rated significantly higher than World for Visual in Semi-Stationary ($p_{\mathrm{Holm}} = .008$, $|r_{\mathrm{rb}}| = .83$) and for Key and Visual in Moving (Key: $p_{\mathrm{Holm}} = .002$, $|r_{\mathrm{rb}}| = .93$; Visual: $p_{\mathrm{Holm}} < .001$, $|r_{\mathrm{rb}}| = 1.00$).

While mobility affected whether world- or body-anchored interfaces received higher usability ratings, ratings of the individual body anchors varied largely across interface types in the Semi-Stationary and Moving conditions.
During Moving, Torso received median ratings of 6.0 for both Key ($IQR = 5.5$-$7.0$) and Visual ($IQR = 6.0$-$7.0$).
Head also received a median of 6.0 for Key ($IQR = 5.0$-$7.0$), while its median for Visual was 5.0 ($IQR = 4.0$-$6.5$).


We hypothesise this has to do with the interaction frequency of the controls interface, which participants could quickly bring to perform brief, one-off interactions, whereas the other interfaces required more sustained visual interaction.
We explore participants' reasoning further in Section ~\ref{sec:reasoning}.

\subsection{Anchoring Usage Patterns}

Next, we move from analyzing participants stated preferences, to investigating their actual behavior during the experiment. 

We observed that participants mostly stuck with their initial configuration of anchoring choices before they started the seven  experimental trials.
Only in 6 out of 57 instances (19 participants over 3 mobility conditions each) did participants re-adjusted their choice during these trials. 
In other words, once participants found a configuration that worked well for them during the practice trials, they rarely changed their anchoring configuration during an experiment.

To better understand how participants configured the interface types, and compare this to their stated preferences, we investigate for each mobility condition, how much anchoring choices varied across the different interface types.
Therefore, we calculate the number of \emph{anchor differences} across interface types by counting the number of different anchoring choices within a condition and subtracting one (using the final anchoring configuration). 
For instance, if the same anchoring mode (e.g., torso) is used consistently across the three interface types, there are 0 anchor differences. 
If all three interface types use a different anchor (e.g., key - world; visual - torso; controls - arm) we count 2 anchor differences, the maximum possible in our experiments.
In the Stationary condition, participants mostly used the same anchor across the three interface types (median $=0$, $IQR = 0$-$0$), with 84\% choosing only World anchoring.
This pattern shifted substantially in the Semi-Stationary and Moving conditions, where the number of anchor differences had a median of 1 ($IQR = 1$-$2$) in both conditions.
Participants used different anchors across the three interface types in 84\% of Semi-Stationary cases and 79\% of Moving cases, compared with 16\% of Stationary cases.


The increased difference in anchoring settings across interface types in conditions involving user locomotion reflects the emergence of task-specific anchoring strategies.
For instance, during the study participants maintained a consistent anchor for key and visual tasks (typically Torso or Head) but mostly used arm anchoring for controls.

\subsection{Reasoning for Anchoring Preferences}
\label{sec:reasoning}
Our thematic analysis of open-ended questions resulted in six related but analytically distinct themes.
Because a single explanation could express multiple rationales, responses received multiple theme labels; the themes are therefore analytically distinct but not mutually exclusive. For clarity, each quotation below is discussed under the theme most central to the excerpt.
For each quotation, we indicate the participant ID, mobility condition, and anchoring mode using the format \textit{(ID / mobility / anchoring)}.

\textbf{Theme 1: Immediate Access} was the most prominent theme.
This theme captures whether participants could quickly see, reach, or interact with an interface when needed.
It concerns the immediate availability of the interface, rather than how movement affected the suitability of its reference frame, whether it obstructed the FoV, or why it was needed.
In stationary contexts, participants associated access with immediate visibility and reachability, noting, for example, ``I can reach and see [all UI elements when I need] \textit{(P3 / stat. / world)}''.
Participants also described how body-relative anchors could provide immediate access: ``[With torso anchoring the interface] is easily accessible \textit{(P19 / moving / torso)}''.
Similar reasoning supported other anchoring choices, as participants noted ``It's easier to access the controls by just flipping my hand \textit{(P5 / semi-stat. / arm)}'', and ``[using head anchoring], I have the controls always accessible. If they are well positioned, I think it's [...] practical and fast. \textit{(P15 / moving / head)}''.
These comments focus on the resulting availability of the interface.

\textbf{Theme 2: Movement Context} emerged as the second most frequently mentioned consideration.
This theme captures participants' explicit reasoning about being stationary or moving and whether an environmental or body-relative reference frame remained appropriate under that movement.
In stationary contexts, participants referred directly to the absence of movement to justify World anchoring: ``I did not need to move \textit{(P1 / stat. / world)}'', and ``I was stationary [... world anchoring was more stable] \textit{(P4 / stat. / world)}''.
Under locomotion, participants emphasized how body anchoring preserved a consistent spatial relationship to themselves: ``With arm [anchoring] I bring it with me \textit{(P17 / moving / arm)}''.
These comments show that movement changed which reference frame participants considered appropriate: world anchoring was associated with stationary use, whereas body-relative anchoring maintained a relationship to the user during locomotion.

\textbf{Theme 3: Predictability \& Mental Model} emerged as a recurring theme in participants' reasoning, reflecting the importance of anticipating where an interface would be located and how it would behave.
Participants emphasized that forming a reliable mental map of the UI was critical for quickly locating each interface.
In stationary settings, predictability was associated with consistent spatial placement, with participants noting that ``I'd rather know where I need to look to see the UI I need \textit{(P2 / stat. / world)}''.
Participants also valued anchors that preserved a known interface location as their position changed, noting that ``Knowing that the UI is exactly in that place is important for me \textit{(P15 / semi-stat. / head)}''.
Some participants used body-relative metaphors to support their mental models.
For example, as a reason for selecting arm anchoring: ``Arm had good intuition as if I had a watch. It was better in terms of organization to know it was in my arm \textit{(P11 / moving / arm)}''.

\textbf{Theme 4: FoV Management \& Visual Clutter} captures how interfaces occupied or obstructed the field of view and competed with other visual content.
In stationary settings, participants associated World anchoring with reduced visual interference, noting that ``[Using world anchoring] there was no clutter in my FoV \textit{(P15 / stat. / world)}''.
When moving, concerns about visual clutter became more pronounced, particularly in relation to head anchoring.
Participants justified a lower preference for head anchoring by stating that ``I don't like to have [UIs constantly] in my FoV so I did not like head [anchoring] as much \textit{(P14 / moving / head)}''.
Some participants selected torso and arm anchoring to keep interfaces outside the FoV when they were not being used: ``Arm allowed me to quickly access the controls without having it always in my FoV \textit{(P18 / semi-stat. / arm)}'', and ``[Torso anchoring] would not take space from my FoV \textit{(P12 / semi-stat. / torso)}''.
Participants also indicated that tolerance for FoV positioning depended on interface size.
Smaller interfaces were acceptable even when head anchored, as reflected in the comment ``[Controls] were small so it would not obstruct my FoV too much, therefore it was fine to have them head anchored \textit{(P4 / moving / head)}''.

\textbf{Theme 5: Task Characteristics} captures how anchoring choices were matched to the functional characteristics of an interface, including its priority, urgency, interaction frequency, and whether it was location-bound.
High-priority or time-critical interfaces were described as tolerating more visual intrusion in exchange for guaranteed availability: ``More urgent tasks have to be easily accessible so head [anchoring] was nice \textit{(P16 / moving / head)}''.
Location-bound activities motivated the use of World anchoring, when associated with two task locations: ``It was more practical to have it anchored to world as the task and information I needed was all in the same place \textit{(P12 / semi-stat. / world)}''.

Interaction frequency also shaped anchoring choices.
Lower-priority and intermittently used interfaces were often assigned body-relative anchors that allowed them to remain unobtrusive between interactions: ``As a secondary task [(visual)] I would only use it when I needed by moving the arm \textit{(P10 / moving / arm)}''.

\textbf{Theme 6: Interaction Stability} affected usability, particularly in the context of hand-based interaction.
Participants frequently highlighted how the relative motion of UI elements with respect to their hands influenced interaction accuracy and effort.
A recurring concern was that head and torso anchoring could introduce unintended UI motion due to natural, involuntary body movements.
Participants reported that even small, unintended head or torso shifts caused the interface to jitter or drift relative to the hand, making input more difficult.
As one participant noted, ``With head [anchoring] the interface would be moving a lot \textit{(P12 / stat. / world)}'', while another stated that ``[I don't always] move my torso intentionally \textit{(P16 / semi-stat / torso)}''.
As a result, anchoring methods that provided greater stability relative to the hand were often preferred. 
World anchoring was frequently described as inherently stable (``World was the most stable \textit{(P15 / stat. / world)}''), while arm anchoring was valued for enabling higher stability, as mentioned: ``Arm [anchoring made] it easier to interact because the [UI] was very stable \textit{(P3 / moving / arm)}''.
One participant emphasized how ``Having [the UI] still made it easier [to interact] as the interfaces would not mix up with each other \textit{(P13 / stat. / world)}'', pointing up the the possible inconvenience of having UIs overlaping one another when using different anchoring targets.

\section{Discussion}
The goal of our study was to understand how users prefer to anchor MR UIs across different mobility settings and interfaces and their reasoning behind these preferences, in order to draw design implications from them.
Based on the quantitative and qualitative results presented in the previous section, we synthesize how locomotion, task characteristics, and UI properties shape anchoring behavior, and discuss what these findings imply for the design of MR interfaces.

\subsection{Anchoring Preferences Across Usage Contexts}
\textbf{Locomotion:} across our analyses, locomotion emerged as the main determinant of whether users prefered world or body-anchored interfaces. 
When stationary, participants showed an overwhelming and highly confident preference for world anchoring across all tasks, reflected not only in near-unanimous usage rates (94.7–100\%) but also in the usability ratings.
This confirms from a subjective-preference perspective what prior performance-oriented studies have shown for stability and precision: target acquisition and cognitive load improve as content becomes more stable relative to the physical environment \cite{ghasemi2021headlocked, tabone2024insights}, along with predictability from a coherent spatial mental model of the UI.

When the scenario starts involving locomotion, this preference structure changes.
Declared preferences and usability ratings show a sharp decline in the suitability of world anchoring under movement, with ratings dropping to low usability in moving conditions.
This shift is consistent with prior work showing that egocentric, different body-anchored frames become preferable during movement as they reduce the need for visual reorientation \cite{jannat2025handUIs, rasch2025track}. 
Our results extend this literature by showing that this shift is not merely a performance effect but a conscious preference.
Furthermore, they highlight the importance of supporting individual body anchors (such as head, torso, and arms), instead of treating it as a single category \cite{pei2024mobility}.

The advantages of world anchoring are no longer relevant as movement increases and body-anchored interfaces are preferred as they maintain spatial coherence relative to the user, preserving access and usability as the user moves.
Intermediate mobility levels illustrate this transition, aligning with the mixed and sometimes contradictory results reported across prior anchoring studies \cite{manakhov2024gaze, rasch2025track}, which likely originate from differences across experimental designs.
This suggests that moderate mobility opens up a design space in which multiple anchoring strategies can be viable, depending on task demands.
Notably, 42\% of the participants still reported a world-anchoring preference for the key interface type in semi-stationary settings, reflecting deliberate planning of UI layouts around a task \cite{jo2023flowar, acherki2025arsurgery}, leveraging the stability of world anchoring when UI supports a location-bound actity.

\textbf{Task:} 
In conditions involving movement, the interface type primarily influenced \emph{which} body anchor was preferred. Because the three interfaces jointly differed in several properties, these results should not be interpreted as isolating the effect of any single property.
For example, in the moving condition, body anchoring was preferred across all tasks but no single anchor prevailed for all the UIs.
For instance, participants noted that tasks with high urgency such as picking up a call led them to select anchoring options that make the UI easily available (an explanation stated for selecting head anchoring).
Another point that came across in participant's comments was that the frequency which they had to interact with the UI was also a factor in their anchoring decision. 
UIs requiring constant visual and hand input, such as the key interface, led participants to select anchoring modes they deemed more accessible, typically within a quickly accessible viewing and reachable region at the cost of potential clutter \cite{ghasemi2021headlocked, mack2023headplacements}.
In contrast, sporadic tasks such as picking up a call in the controls UI, had users placing them in anchoring modes that would allow access on-demand.
For example, arm anchoring was frequently used for this interface as participants reported that it would make the UI accessible when they deemed it necessary with a hand gesture, matching earlier evidence that arm-anchored interfaces support comfortable, low-effort repeated access \cite{jannat2025handUIs}.
While our results support previous findings, they reveal an important insight that previous studies limited to evaluating a single anchor \cite{jannat2025handUIs} or comparing multiple anchors for the same task \cite{manakhov2024gaze, rasch2025track} could not establish: no anchor strategy is universally superior across interface types and mobility settings. 
Instead, participants balance trade-offs between accessibility, spatial coherence, and attentional overhead according to the functional role and characteristics of each UI element within a given task.

\textbf{UI Properties:} size and interaction modality of the UIs were factors that participants consistently mentioned that shaped their anchoring decisions.
They consistently reported greater tolerance for smaller interfaces appearing in their FoV, making head anchoring acceptable for compact UIs.
As UI size increases, preferences shifted towards anchors that mitigated visual clutter, such as torso, arm, or world anchoring.
This is consistent with prior findings showing how Head-anchoring is a source of FoV interference and cognitive load during visually demanding tasks \cite{ghasemi2021headlocked, mack2023headplacements, rasch2025track}.
Interaction modality also played a critical role. 
For interfaces requiring precise hand interaction, stability relative to the hand became a dominant concern.
Participants repeatedly noted that involuntary head or torso motion introduced jitter when using Head or Torso anchoring, negatively affecting hand interaction accuracy.
World anchoring and arm anchoring were perceived as more stable in this regard, either by fixing the interface in space or by coupling it directly to the hand.
These findings help explain why Arm anchoring was strongly preferred for the Controls task.

\subsection{User personalization and implications for design}
Our results point to two main findings regarding how users make anchoring decisions and how these decisions should inform MR UI design.
First, while we observe clear high-level patterns - most notably the shift from world-anchored to body anchored interfaces as users transition from stationary to moving contexts - we do not find a single body anchor (e.g., head, torso, or arm) that is universally preferred for a given interface type.
Although certain anchors, such as torso, achieved higher usability ratings in most cases during locomotion, participants would select different anchoring options in the study and justified them based on personal preferences, individual perceptions of task demands, or perceived comfort.
This variability indicates that anchoring effectiveness is highly individual and context dependent, aligning with existing findings showing how personalized MR configurations vary substantially across individuals even for similar tasks \cite{cho2024minexr}.
Although some of this uncertainty can be addressed automatically \cite{cheng2021semanticadapt, cheng2023interactionadapt, zhipeng2024situationadapt}, MR systems should prioritize user choice and support diverse anchoring configurations, rather than forcing a single universal strategy.

Second, once users identified an anchoring configuration that supported their task, they would rarely change it.
Despite being explicitly encouraged to experiment with multiple anchoring configurations during the first three practice trials, participants converged on a preferred setup early and maintained it throughout the task.
This behavior suggests that anchoring is perceived as a configuration decision that is unlikely to adjust frequently if context stays the same.
Consequently, the initial onboarding or setup phase is critical for the usability of MR UIs.
In realistic settings, users may not revisit anchoring settings after this initial stage, even if their initial configuration is suboptimal.
These findings highlight the importance of designing MR systems that facilitate the discovery of effective anchoring configurations during first use.
This may include guided setup processes, structured exploration of anchoring options, or tutorials that explicitly expose users to the trade-offs between different anchors.

\subsection{Limitations}
Given the exploratory design and sample size (N=19), the quantitative findings should be interpreted as convergent evidence alongside participants’ declared preferences, observed anchoring behavior, and qualitative accounts rather than as definitive population-level estimates.
Although we intentionally designed three interfaces with distinct interaction requirements, UI design in MR involves many interdependent factors that are difficult to isolate experimentally.
Properties such as interface size, spatial extent, interaction technique, frequency of use, and whether a task is location-bound are inherently intertwined and all influence user preferences for anchoring choices.
Therefore, some of the observed anchoring preferences may reflect combined effects of these factors rather than the influence of the anchoring alone.
This complexity reflects the reality of MR interface design and limits the extent to which we can make strong causal claims from individual quantitative comparisons.
In our study, the number and diversity of interface types was necessarily limited.
While the selected interfaces were representative of compelling MR use cases, it is not possible to cover the whole design space of MR UIs within a single study.
This is a general challenge for research in this area and motivates our emphasis on qualitative analyses.
By focusing on participants' reasoning rather than performance-based measures, we get insights that extend beyond the specific interfaces evaluated.
Finally, interfaces were world anchored at the beginning of each condition, resembling standard configurations in most MR applications. Participants were explicitly asked  to configure each interface before beginning the seven analyzed trials, reducing the risk of the default configuration being evaluated as a preference.

\subsection{Future Work}
The findings of our study suggest promising directions for future research on anchoring strategies for MR UIs.
A common theme emerging from our results is the trade-off between stability and accessibility across mobility contexts.
While world anchoring was consistently perceived as the most stable and easy to interact option, its usability worsened substantially in conditions involving locomotion.
This points to opportunities for adaptive anchoring that dynamically transition between world- and body- anchored modes based on user movement, task, or interaction intent. 
Future work could explore how to implement such transitions without disrupting spatial continuity or the user's mental model of the UI.

Another recurring issue was visual clutter and attentional load, particularly when multiple interfaces are anchored to the body during movement.
Future work could investigate context-aware UI adaptations \cite{lindlbauer2019context}, such as dynamically adjusting opacity or level of detail based on interaction frequency and gaze behavior.
Interfaces might also temporarily collapse, minimize, or change position when not in use, reappearing only when the system detects user intent.

While this work focused on anchoring choices at the level of discrete anchor types, our data revealed differences in how users spatially position UIs within a given anchoring mode (e.g., relative offsets, preferred distances, or position in the FoV). 
Follow up research could analyze these positioning strategies and their relationship to ergonomics \cite{belo2021xrgonomics}, reachability, and performance.
Such insights could inform default placement suggestions or personalized layout recommendations within different anchoring modes.

Finally, our work examined anchoring decisions within relatively short-term usage scenarios. Longitudinal studies can reveal how anchoring preferences change with prolonged use, increased familiarity, or changing tasks and environments.
This includes investigating whether users are willing to adjust anchoring configurations over time, or whether early configurations stay prevalent.

\section{Conclusion}

In this work, we examined how mobility and UI properties shape anchoring preferences through a user study in which participants freely configured anchoring modes across multiple interface types.
Our findings show that locomotion is the primary factor influencing anchoring preference.
When stationary, users strongly prefer world-anchored interfaces due to their stability and predictability.
As movement increases, preferences shift toward body-relative anchors, which better maintain accessibility and spatial coherence.
No single body anchor dominated across interfaces.
Instead, users selected anchors based on interface characteristics and personal preference, highlighting the contextual and individualized nature of anchoring strategies.
Once users found a configuration that worked well, they rarely changed it during later interactions.

These results suggest that MR systems should prioritize user customization and flexible anchoring rather than relying on a single default strategy.
The onboarding phase is particularly important, as initial anchoring choices tend to persist.
By understanding how mobility, task demands, and UI properties affect anchoring preferences, our work can help guide the design of more adaptable and usable MR systems.

\bibliographystyle{abbrv-doi}

\bibliography{references}

\appendix
\clearpage
\onecolumn

\section{Study Details}
\label{app:details}
Here we list the questionnaire items (\textbf{Q}) and interview prompts (\textbf{P}) asked after each mobility condition (three times, one for Stationary, one for Semi-Stationary, and one for Moving).

\begin{itemize}
    \item \textbf{Q:} Order the Anchoring modes according to your preference for the Key instruction interface (Options: World / Head / Torso / Arm).
    \item \textbf{P:} Explain your decision.
     \item \textbf{Q:} For the Key instruction interface, answer the following question, from 1 (strongly disagree) to 7 (strongly agree): Using the following anchoring mode was/would make the interface easy to use. (Participants gave an answer in a Likert scale for each anchoring mode: World / Head / Torso / Arm)
     \item \textbf{Q:} For the Key instruction interface, answer the following question, from 1 (strongly disagree) to 7 (strongly agree). Using the following anchoring mode I had to spend too much time correcting things. (Participants gave an answer in a Likert scale for each anchoring mode: World / Head / Torso / Arm)
     \item \textbf{Q:} Order the Anchoring modes according to your preference for the Visual interface (Options: World / Head / Torso / Arm).
    \item \textbf{P:} Explain your decision.
     \item \textbf{Q:} For the Visual interface, answer the following question, from 1 (strongly disagree) to 7 (strongly agree): Using the following anchoring mode was/would make the interface easy to use. (Participants gave an answer in a Likert scale for each anchoring mode: World / Head / Torso / Arm)
     \item \textbf{Q:} For the Visual interface, answer the following question, from 1 (strongly disagree) to 7 (strongly agree). Using the following anchoring mode I had to spend too much time correcting things. (Participants gave an answer in a Likert scale for each anchoring mode: World / Head / Torso / Arm)
     \item \textbf{Q:} Order the Anchoring modes according to your preference for the Controls interface (Options: World / Head / Torso / Arm).
     \item \textbf{P:} Explain your decision.
     \item \textbf{Q:} For the Controls instruction interface, answer the following question, from 1 (strongly disagree) to 7 (strongly agree): Using the following anchoring mode was/would make the interface easy to use. (Participants gave an answer in a Likert scale for each anchoring mode: World / Head / Torso / Arm)
     \item \textbf{Q:} For the Controls instruction interface, answer the following question, from 1 (strongly disagree) to 7 (strongly agree). Using the following anchoring mode I had to spend too much time correcting things. (Participants gave an answer in a Likert scale for each anchoring mode: World / Head / Torso / Arm)
     \item \textbf{P:} Any further thoughts on preferences? Did you change your mind since the planning phase?
     \item \textbf{P:} What could be improved for the anchoring / positioning to work better in this scenario?
     \item \textbf{P:} Other Comments
\end{itemize}

The study was performed in a large and quiet room of 6x4.5 meters. For the Stationary condition, participants completed the tasks while sitting in a chair. For the Semi-stationary there were two working zones in opposite zones of the room (around 5m distance. For the Moving condition, the layout in Figure \ref{fig:teaser} (moving) occupied the whole room.

The interface panels had the following approximate dimensions: Key Task - UI 24cm (height) * 15cm (width); Visual Task UI - 20cm (height) * 15cm (width); Control Panel - 6cm (height) * 7cm (width).

\section{Full Statistical Results}
\label{app:full-statistics}

This appendix reports the complete inferential analyses of the
ease-of-use ratings.
All analyses used repeated measures from the same 19 participants.
Friedman tests were used for omnibus comparisons, followed by
paired Wilcoxon signed-rank tests when the omnibus result was
significant.
Pairwise $p$-values were corrected using the Holm procedure.
Kendall's $W$ is reported for omnibus effects, and the absolute
rank-biserial correlation, $|r_{\mathrm{rb}}|$, is reported for
pairwise effects.
Bold $p$-values indicate statistical significance at $\alpha=.05$.
Significance decisions were based on unrounded values.

\subsection{Comparisons Among Anchoring Methods}
\label{app:anchor-comparisons}

Table~\ref{tab:supp-anchor-omnibus} reports comparisons among
World, Head, Torso, and Arm anchoring within each mobility
condition and interface type.
Table~\ref{tab:supp-anchor-posthoc} reports all six pairwise
anchor comparisons within every mobility-condition
$\times$ interface-type combination.


\small
\setlength{\LTleft}{\fill}
\setlength{\LTright}{\fill}
\setlength{\LTcapwidth}{\textwidth}
\setlength{\tabcolsep}{7pt}
\renewcommand{\arraystretch}{1.12}

\setlength{\LTleft}{\fill}
\setlength{\LTright}{\fill}
\setlength{\LTcapwidth}{\textwidth}
\begin{longtable}{@{}llccc@{}}
\caption{Omnibus Friedman tests comparing the four anchoring
methods within each mobility-condition $\times$ interface-type
combination ($n=19$, $df=3$).}
\label{tab:supp-anchor-omnibus}\\

\toprule
\textbf{Mobility} &
\textbf{Interface} &
$\boldsymbol{\chi^2}$ &
\textbf{$p$} &
\textbf{Kendall's $W$} \\
\midrule
\endfirsthead

\multicolumn{5}{c}{
\textbf{Table~\thetable{} continued from the previous page}}\\
\toprule
\textbf{Mobility} &
\textbf{Interface} &
$\boldsymbol{\chi^2}$ &
\textbf{$p$} &
\textbf{Kendall's $W$} \\
\midrule
\endhead

\midrule
\multicolumn{5}{r}{Continued on the next page}\\
\endfoot

\bottomrule
\endlastfoot

Stationary      & Key      & 40.72 & \textbf{$7.48\times10^{-9}$} & 0.71 \\
Stationary      & Visual   & 40.42 & \textbf{$8.69\times10^{-9}$} & 0.71 \\
Stationary      & Controls & 31.52 & \textbf{$6.62\times10^{-7}$} & 0.55 \\
\midrule
Semi-Stationary & Key      & 9.44  & \textbf{.024}                 & 0.17 \\
Semi-Stationary & Visual   & 17.53 & \textbf{$5.50\times10^{-4}$} & 0.31 \\
Semi-Stationary & Controls & 17.42 & \textbf{$5.78\times10^{-4}$} & 0.31 \\
\midrule
Moving          & Key      & 24.32 & \textbf{$2.14\times10^{-5}$} & 0.43 \\
Moving          & Visual   & 29.26 & \textbf{$1.98\times10^{-6}$} & 0.51 \\
Moving          & Controls & 25.55 & \textbf{$1.18\times10^{-5}$} & 0.45 \\

\end{longtable}


\begin{longtable}{
  @{}
  L{2.5cm}
  L{1.8cm}
  L{3.5cm}
  C{2.4cm}
  C{1.8cm}
  L{3cm}
  @{}
}
\caption{Holm-corrected paired Wilcoxon signed-rank comparisons
between anchoring methods within each mobility-condition
$\times$ interface-type combination ($n=19$).
The higher-rated anchor is reported only for statistically
significant contrasts. Effect sizes are absolute rank-biserial
correlations.}
\label{tab:supp-anchor-posthoc}\\

\toprule
\textbf{Mobility} &
\textbf{Interface} &
\textbf{Contrast} &
$\boldsymbol{p_{\mathrm{Holm}}}$ &
$\boldsymbol{|r_{\mathrm{rb}}|}$ &
\textbf{Higher-rated anchor} \\
\midrule
\endfirsthead

\multicolumn{6}{c}{
\textbf{Table~\thetable{} continued from the previous page}}\\
\toprule
\textbf{Mobility} &
\textbf{Interface} &
\textbf{Contrast} &
$\boldsymbol{p_{\mathrm{Holm}}}$ &
$\boldsymbol{|r_{\mathrm{rb}}|}$ &
\textbf{Higher-rated anchor} \\
\midrule
\endhead

\midrule
\multicolumn{6}{r}{Continued on the next page}\\
\endfoot

\bottomrule
\endlastfoot

Stationary & Key & World vs.\ Head
  & \textbf{$6.39\times10^{-4}$} & 1.00 & World $>$ Head \\
Stationary & Key & World vs.\ Torso
  & \textbf{$7.09\times10^{-4}$} & 1.00 & World $>$ Torso \\
Stationary & Key & World vs.\ Arm
  & \textbf{$6.39\times10^{-4}$} & 1.00 & World $>$ Arm \\
Stationary & Key & Head vs.\ Torso
  & .855 & 0.05 & --- \\
Stationary & Key & Head vs.\ Arm
  & \textbf{.038} & 0.66 & Head $>$ Arm \\
Stationary & Key & Torso vs.\ Arm
  & \textbf{.022} & 0.73 & Torso $>$ Arm \\

\midrule

Stationary & Visual & World vs.\ Head
  & \textbf{$6.42\times10^{-4}$} & 1.00 & World $>$ Head \\
Stationary & Visual & World vs.\ Torso
  & \textbf{.001} & 1.00 & World $>$ Torso \\
Stationary & Visual & World vs.\ Arm
  & \textbf{$6.42\times10^{-4}$} & 1.00 & World $>$ Arm \\
Stationary & Visual & Head vs.\ Torso
  & .620 & 0.14 & --- \\
Stationary & Visual & Head vs.\ Arm
  & .106 & 0.54 & --- \\
Stationary & Visual & Torso vs.\ Arm
  & \textbf{.021} & 0.74 & Torso $>$ Arm \\

\midrule

Stationary & Controls & World vs.\ Head
  & \textbf{.001} & 1.00 & World $>$ Head \\
Stationary & Controls & World vs.\ Torso
  & \textbf{.001} & 1.00 & World $>$ Torso \\
Stationary & Controls & World vs.\ Arm
  & \textbf{.001} & 0.97 & World $>$ Arm \\
Stationary & Controls & Head vs.\ Torso
  & .565 & 0.16 & --- \\
Stationary & Controls & Head vs.\ Arm
  & .447 & 0.33 & --- \\
Stationary & Controls & Torso vs.\ Arm
  & .142 & 0.54 & --- \\

\midrule

Semi-Stationary & Key & World vs.\ Head
  & 1.000 & 0.03 & --- \\
Semi-Stationary & Key & World vs.\ Torso
  & 1.000 & 0.15 & --- \\
Semi-Stationary & Key & World vs.\ Arm
  & .750 & 0.34 & --- \\
Semi-Stationary & Key & Head vs.\ Torso
  & .750 & 0.34 & --- \\
Semi-Stationary & Key & Head vs.\ Arm
  & .432 & 0.47 & --- \\
Semi-Stationary & Key & Torso vs.\ Arm
  & .159 & 0.59 & --- \\

\midrule

Semi-Stationary & Visual & World vs.\ Head
  & \textbf{.039} & 0.67 & Head $>$ World \\
Semi-Stationary & Visual & World vs.\ Torso
  & \textbf{.008} & 0.83 & Torso $>$ World \\
Semi-Stationary & Visual & World vs.\ Arm
  & .521 & 0.31 & --- \\
Semi-Stationary & Visual & Head vs.\ Torso
  & .521 & 0.31 & --- \\
Semi-Stationary & Visual & Head vs.\ Arm
  & .204 & 0.47 & --- \\
Semi-Stationary & Visual & Torso vs.\ Arm
  & \textbf{.008} & 0.81 & Torso $>$ Arm \\

\midrule

Semi-Stationary & Controls & World vs.\ Head
  & \textbf{.038} & 0.67 & Head $>$ World \\
Semi-Stationary & Controls & World vs.\ Torso
  & \textbf{.005} & 0.92 & Torso $>$ World \\
Semi-Stationary & Controls & World vs.\ Arm
  & \textbf{.018} & 0.76 & Arm $>$ World \\
Semi-Stationary & Controls & Head vs.\ Torso
  & .605 & 0.38 & --- \\
Semi-Stationary & Controls & Head vs.\ Arm
  & .940 & 0.19 & --- \\
Semi-Stationary & Controls & Torso vs.\ Arm
  & .968 & 0.01 & --- \\

\midrule

Moving & Key & World vs.\ Head
  & \textbf{.013} & 0.80 & Head $>$ World \\
Moving & Key & World vs.\ Torso
  & \textbf{.002} & 0.93 & Torso $>$ World \\
Moving & Key & World vs.\ Arm
  & .057 & 0.57 & --- \\
Moving & Key & Head vs.\ Torso
  & .196 & 0.36 & --- \\
Moving & Key & Head vs.\ Arm
  & \textbf{.050} & 0.64 & Head $>$ Arm \\
Moving & Key & Torso vs.\ Arm
  & \textbf{.013} & 0.77 & Torso $>$ Arm \\

\midrule

Moving & Visual & World vs.\ Head
  & \textbf{.011} & 0.82 & Head $>$ World \\
Moving & Visual & World vs.\ Torso
  & \textbf{$7.42\times10^{-4}$} & 1.00 & Torso $>$ World \\
Moving & Visual & World vs.\ Arm
  & \textbf{.027} & 0.70 & Arm $>$ World \\
Moving & Visual & Head vs.\ Torso
  & \textbf{.031} & 0.64 & Torso $>$ Head \\
Moving & Visual & Head vs.\ Arm
  & .371 & 0.23 & --- \\
Moving & Visual & Torso vs.\ Arm
  & \textbf{.022} & 0.74 & Torso $>$ Arm \\

\midrule

Moving & Controls & World vs.\ Head
  & \textbf{.007} & 0.86 & Head $>$ World \\
Moving & Controls & World vs.\ Torso
  & \textbf{.001} & 0.96 & Torso $>$ World \\
Moving & Controls & World vs.\ Arm
  & \textbf{.002} & 0.92 & Arm $>$ World \\
Moving & Controls & Head vs.\ Torso
  & .631 & 0.37 & --- \\
Moving & Controls & Head vs.\ Arm
  & .631 & 0.30 & --- \\
Moving & Controls & Torso vs.\ Arm
  & .631 & 0.20 & --- \\

\end{longtable}

\subsection{Comparisons Among Mobility Conditions}
\label{app:mobility-comparisons}

Table~\ref{tab:supp-mobility-omnibus} reports comparisons across
the Stationary, Semi-Stationary, and Moving conditions for every
anchoring-method $\times$ interface-type combination.
Table~\ref{tab:supp-mobility-posthoc} reports all three pairwise
mobility comparisons for each combination.


\begin{longtable}{@{}llccc@{}}
\caption{Omnibus Friedman tests comparing the three mobility
conditions within each anchoring-method $\times$ interface-type
combination ($n=19$, $df=2$).}
\label{tab:supp-mobility-omnibus}\\

\toprule
\textbf{Interface} &
\textbf{Anchor} &
$\boldsymbol{\chi^2}$ &
\textbf{$p$} &
\textbf{Kendall's $W$} \\
\midrule
\endfirsthead

\multicolumn{5}{c}{
\textbf{Table~\thetable{} continued from the previous page}}\\
\toprule
\textbf{Interface} &
\textbf{Anchor} &
$\boldsymbol{\chi^2}$ &
\textbf{$p$} &
\textbf{Kendall's $W$} \\
\midrule
\endhead

\midrule
\multicolumn{5}{r}{Continued on the next page}\\
\endfoot

\bottomrule
\endlastfoot

Key      & World & 28.56 & \textbf{$6.29\times10^{-7}$} & 0.75 \\
Key      & Head  & 17.09 & \textbf{$1.94\times10^{-4}$} & 0.45 \\
Key      & Torso & 21.72 & \textbf{$1.92\times10^{-5}$} & 0.57 \\
Key      & Arm   & 6.53  & \textbf{.038}                 & 0.17 \\
\midrule
Visual   & World & 32.38 & \textbf{$9.32\times10^{-8}$} & 0.85 \\
Visual   & Head  & 9.59  & \textbf{.008}                 & 0.25 \\
Visual   & Torso & 18.14 & \textbf{$1.15\times10^{-4}$} & 0.48 \\
Visual   & Arm   & 7.24  & \textbf{.027}                 & 0.19 \\
\midrule
Controls & World & 31.73 & \textbf{$1.29\times10^{-7}$} & 0.83 \\
Controls & Head  & 10.65 & \textbf{.005}                 & 0.28 \\
Controls & Torso & 7.59  & \textbf{.022}                 & 0.20 \\
Controls & Arm   & 17.75 & \textbf{$1.40\times10^{-4}$} & 0.47 \\

\end{longtable}


\begin{longtable}{
  @{}
  L{1.8cm}
  L{1.8cm}
  L{5.0cm}
  C{2.4cm}
  C{1.8cm}
  L{3cm}
  @{}
}
\caption{Holm-corrected paired Wilcoxon signed-rank comparisons
between mobility conditions for each anchoring-method
$\times$ interface-type combination ($n=19$).
The higher-rated condition is reported only for statistically
significant contrasts. Effect sizes are absolute rank-biserial
correlations.}
\label{tab:supp-mobility-posthoc}\\

\toprule
\textbf{Interface} &
\textbf{Anchor} &
\textbf{Contrast} &
$\boldsymbol{p_{\mathrm{Holm}}}$ &
$\boldsymbol{|r_{\mathrm{rb}}|}$ &
\textbf{Higher-rated condition} \\
\midrule
\endfirsthead

\multicolumn{6}{c}{
\textbf{Table~\thetable{} continued from the previous page}}\\
\toprule
\textbf{Interface} &
\textbf{Anchor} &
\textbf{Contrast} &
$\boldsymbol{p_{\mathrm{Holm}}}$ &
$\boldsymbol{|r_{\mathrm{rb}}|}$ &
\textbf{Higher-rated condition} \\
\midrule
\endhead

\midrule
\multicolumn{6}{r}{Continued on the next page}\\
\endfoot

\bottomrule
\endlastfoot

Key & World & Stationary vs.\ Semi-Stationary
  & \textbf{.003} & 1.00 & Stationary $>$ Semi-Stationary \\
Key & World & Stationary vs.\ Moving
  & \textbf{$4.32\times10^{-4}$} & 1.00 & Stationary $>$ Moving \\
Key & World & Semi-Stationary vs.\ Moving
  & \textbf{.003} & 0.97 & Semi-Stationary $>$ Moving \\

\midrule

Key & Head & Stationary vs.\ Semi-Stationary
  & \textbf{.009} & 0.91 & Semi-Stationary $>$ Stationary \\
Key & Head & Stationary vs.\ Moving
  & \textbf{.003} & 0.94 & Moving $>$ Stationary \\
Key & Head & Semi-Stationary vs.\ Moving
  & .077 & 0.62 & --- \\

\midrule

Key & Torso & Stationary vs.\ Semi-Stationary
  & \textbf{.002} & 0.95 & Semi-Stationary $>$ Stationary \\
Key & Torso & Stationary vs.\ Moving
  & \textbf{.002} & 0.97 & Moving $>$ Stationary \\
Key & Torso & Semi-Stationary vs.\ Moving
  & \textbf{.013} & 0.85 & Moving $>$ Semi-Stationary \\

\midrule

Key & Arm & Stationary vs.\ Semi-Stationary
  & .088 & 0.60 & --- \\
Key & Arm & Stationary vs.\ Moving
  & .088 & 0.63 & --- \\
Key & Arm & Semi-Stationary vs.\ Moving
  & .887 & 0.04 & --- \\

\midrule

Visual & World & Stationary vs.\ Semi-Stationary
  & \textbf{$3.30\times10^{-4}$} & 1.00 & Stationary $>$ Semi-Stationary \\
Visual & World & Stationary vs.\ Moving
  & \textbf{$3.22\times10^{-4}$} & 1.00 & Stationary $>$ Moving \\
Visual & World & Semi-Stationary vs.\ Moving
  & \textbf{.032} & 0.65 & Semi-Stationary $>$ Moving \\

\midrule

Visual & Head & Stationary vs.\ Semi-Stationary
  & \textbf{.021} & 0.78 & Semi-Stationary $>$ Stationary \\
Visual & Head & Stationary vs.\ Moving
  & .107 & 0.56 & --- \\
Visual & Head & Semi-Stationary vs.\ Moving
  & .476 & 0.28 & --- \\

\midrule

Visual & Torso & Stationary vs.\ Semi-Stationary
  & \textbf{.002} & 1.00 & Semi-Stationary $>$ Stationary \\
Visual & Torso & Stationary vs.\ Moving
  & \textbf{.002} & 0.93 & Moving $>$ Stationary \\
Visual & Torso & Semi-Stationary vs.\ Moving
  & .589 & 0.18 & --- \\

\midrule

Visual & Arm & Stationary vs.\ Semi-Stationary
  & .108 & 0.54 & --- \\
Visual & Arm & Stationary vs.\ Moving
  & .108 & 0.58 & --- \\
Visual & Arm & Semi-Stationary vs.\ Moving
  & .154 & 0.47 & --- \\

\midrule

Controls & World & Stationary vs.\ Semi-Stationary
  & \textbf{$3.56\times10^{-4}$} & 1.00 & Stationary $>$ Semi-Stationary \\
Controls & World & Stationary vs.\ Moving
  & \textbf{$3.21\times10^{-4}$} & 1.00 & Stationary $>$ Moving \\
Controls & World & Semi-Stationary vs.\ Moving
  & .120 & 0.55 & --- \\

\midrule

Controls & Head & Stationary vs.\ Semi-Stationary
  & .064 & 0.65 & --- \\
Controls & Head & Stationary vs.\ Moving
  & \textbf{.017} & 0.81 & Moving $>$ Stationary \\
Controls & Head & Semi-Stationary vs.\ Moving
  & .638 & 0.16 & --- \\

\midrule

Controls & Torso & Stationary vs.\ Semi-Stationary
  & \textbf{.033} & 0.69 & Semi-Stationary $>$ Stationary \\
Controls & Torso & Stationary vs.\ Moving
  & \textbf{.033} & 0.70 & Moving $>$ Stationary \\
Controls & Torso & Semi-Stationary vs.\ Moving
  & .819 & 0.07 & --- \\

\midrule

Controls & Arm & Stationary vs.\ Semi-Stationary
  & \textbf{.015} & 0.71 & Semi-Stationary $>$ Stationary \\
Controls & Arm & Stationary vs.\ Moving
  & \textbf{.002} & 0.93 & Moving $>$ Stationary \\
Controls & Arm & Semi-Stationary vs.\ Moving
  & .234 & 0.42 & --- \\

\end{longtable}

\normalsize

\begin{table}[h]
  \centering
  \caption{Median ease-of-use ratings by mobility condition, interface type, and anchoring method. Values in brackets show the interquartile range. Darker green indicates a higher median rating, and bold values indicate the highest median within each interface and mobility combination, including ties. Ratings range from 1 to 7, with higher values indicating greater perceived ease of use.}
  \label{tab:ease-descriptives}
  \scriptsize
  \setlength{\tabcolsep}{1.5pt}
  \renewcommand{\arraystretch}{1.25}
  \begin{tabularx}{\columnwidth}{
    >{\centering\arraybackslash}X
    >{\centering\arraybackslash}X
    >{\centering\arraybackslash}X
    >{\centering\arraybackslash}X
    >{\centering\arraybackslash}X
    >{\centering\arraybackslash}X
  }
    \toprule
    \textbf{Mobility} &
    \textbf{Interface} &
    \textbf{World} &
    \textbf{Head} &
    \textbf{Torso} &
    \textbf{Arm} \\
    \midrule

    \multirow{3}{*}{\textbf{Stationary}}
      & Key
      & \cellcolor{green!55}\shortstack{\textbf{7.0}\\{[7.0--7.0]}}
      & \cellcolor{green!25}\shortstack{5.0\\{[2.0--5.0]}}
      & \cellcolor{green!12}\shortstack{4.0\\{[3.0--5.0]}}
      & \cellcolor{green!2}\shortstack{2.0\\{[1.5--4.0]}} \\

      & Visual
      & \cellcolor{green!55}\shortstack{\textbf{7.0}\\{[7.0--7.0]}}
      & \cellcolor{green!25}\shortstack{5.0\\{[2.0--5.0]}}
      & \cellcolor{green!25}\shortstack{5.0\\{[3.5--5.0]}}
      & \cellcolor{green!5}\shortstack{3.0\\{[2.0--4.0]}} \\

      & Controls
      & \cellcolor{green!55}\shortstack{\textbf{7.0}\\{[7.0--7.0]}}
      & \cellcolor{green!25}\shortstack{5.0\\{[2.0--5.0]}}
      & \cellcolor{green!12}\shortstack{4.0\\{[3.5--5.0]}}
      & \cellcolor{green!5}\shortstack{3.0\\{[2.0--4.0]}} \\

    \midrule

    \multirow{3}{*}{\shortstack{\textbf{Semi-}\\\textbf{Stat.}}}
      & Key
      & \cellcolor{green!25}\shortstack{5.0\\{[3.5--7.0]}}
      & \cellcolor{green!25}\shortstack{5.0\\{[4.0--6.0]}}
      & \cellcolor{green!40}\shortstack{\textbf{6.0}\\{[4.5--6.5]}}
      & \cellcolor{green!12}\shortstack{4.0\\{[3.0--5.0]}} \\

      & Visual
      & \cellcolor{green!2}\shortstack{2.0\\{[2.0--5.0]}}
      & \cellcolor{green!40}\shortstack{6.0\\{[4.0--7.0]}}
      & \cellcolor{green!55}\shortstack{\textbf{7.0}\\{[5.0--7.0]}}
      & \cellcolor{green!12}\shortstack{4.0\\{[2.5--5.0]}} \\

      & Controls
      & \cellcolor{green!2}\shortstack{2.0\\{[1.0--4.0]}}
      & \cellcolor{green!25}\shortstack{5.0\\{[4.0--6.0]}}
      & \cellcolor{green!25}\shortstack{5.0\\{[5.0--6.0]}}
      & \cellcolor{green!55}\shortstack{\textbf{7.0}\\{[3.5--7.0]}} \\

    \midrule

    \multirow{3}{*}{\textbf{Moving}}
      & Key
      & \cellcolor{green!0}\shortstack{1.0\\{[1.0--3.5]}}
      & \cellcolor{green!40}\shortstack{\textbf{6.0}\\{[5.0--7.0]}}
      & \cellcolor{green!40}\shortstack{\textbf{6.0}\\{[5.5--7.0]}}
      & \cellcolor{green!12}\shortstack{4.0\\{[3.0--5.0]}} \\

      & Visual
      & \cellcolor{green!2}\shortstack{2.0\\{[1.0--3.5]}}
      & \cellcolor{green!25}\shortstack{5.0\\{[4.0--6.5]}}
      & \cellcolor{green!40}\shortstack{\textbf{6.0}\\{[6.0--7.0]}}
      & \cellcolor{green!12}\shortstack{4.0\\{[3.0--5.0]}} \\

      & Controls
      & \cellcolor{green!0}\shortstack{1.0\\{[1.0--3.5]}}
      & \cellcolor{green!40}\shortstack{\textbf{6.0}\\{[4.0--6.0]}}
      & \cellcolor{green!40}\shortstack{\textbf{6.0}\\{[4.5--6.0]}}
      & \cellcolor{green!40}\shortstack{\textbf{6.0}\\{[5.0--7.0]}} \\

    \bottomrule
  \end{tabularx}
\end{table}

\end{document}